\documentclass[showpacs,showkeys,preprintnumbers,pre,amsmath,amssymb,superscriptaddress,twocolumn,floatfix]{revtex4-1}

\usepackage[english]{babel}
\usepackage[utf8]{inputenc}
\usepackage[colorinlistoftodos, color=green!40, prependcaption]{todonotes}

\usepackage[colorlinks=true,linkcolor=blue,urlcolor=blue,citecolor=blue]{hyperref}
\usepackage{bm}
\usepackage{mathptmx}
\usepackage{amsmath}

\usepackage{float}
\usepackage{lipsum}

\newcommand{\bxi}{\bm{\xi}}
\newcommand{\bx}{\bm{x}}
\newcommand{\bu}{\bm{u}}
\newcommand{\ea}{\bm{e}_a}

\usepackage{afterpage}

\begin{document}

%%\preprint{APS/123-QED}

\title{Metastable and Unstable Dynamics in multi-phase lattice Boltzmann}

\author{Matteo Lulli} %\email{lulli@sustech.edu.cn}% Your name
\affiliation{Department of Mechanics and Aerospace Engineering, Southern
	University of Science and Technology, Shenzhen, Guangdong 518055, China}
    
\author{Luca Biferale} %\email{biferale@roma2.infn.it}
\affiliation{Department of Physics \& INFN, University of Rome ``Tor Vergata'',
	Via della Ricerca Scientifica 1, 00133 Rome, Italy.}

\author{Giacomo Falcucci} %\email{giacomo.falcucci@uniroma2.it}
\affiliation{Department of Enterprise Engineering ``Mario Lucertini'',
	University of Rome ``Tor Vergata", Via del Politecnico 1, 00133 Rome, Italy;
	John A. Paulson School of Engineering and Applied Physics, {\it Harvard
		University},  33 Oxford Street, 02138 Cambridge, Massachusetts, USA.}

\author{Mauro Sbragaglia} %\email{sbragaglia@roma2.infn.it}
\affiliation{Department of Physics \& INFN, University of Rome ``Tor Vergata'',
	Via della Ricerca Scientifica 1, 00133 Rome, Italy.}

\author{Dong Yang} %\email{yangd3@sustech.edu.cn}
\affiliation{Department of Mechanics and Aerospace Engineering, Southern University of Science and Technology, Shenzhen, Guangdong 518055, China}

\author{Xiaowen Shan}
\email{shanxw@sustech.edu.cn}
\affiliation{Department of Mechanics and Aerospace Engineering, Southern
	University of Science and Technology, Shenzhen, Guangdong 518055, China}

\date{\today}

\begin{abstract}
We quantitatively characterize the metastability in a multi-phase lattice Boltzmann model. The structure factor of density fluctuations is theoretically obtained and numerically verified to a high precision, for all simulated wave-vectors and reduced temperatures. The static structure factor is found to consistently diverge as the temperature approaches the critical-point or the density approaches the spinodal line at a sub-critical temperature. Theoretically predicted critical exponents are observed in both cases. Finally, the phase separation in the unstable branch follows the same pattern, i.e.\ the generation of interfaces with different topology, as observed in molecular dynamics simulations. All results can be independently reproduced through the ``idea.deploy" framework~\href{https://github.com/lullimat/idea.deploy}{\nolinkurl{https://github.com/lullimat/idea.deploy}}
\end{abstract}

\maketitle

Metastability constitutes one of the basic mechanisms for cavitation inception and nucleation in general~\cite{Debenedetti1996,Kalikmanov2013} which are of paramount importance in both fundamental science and critical applications such as sono-luminescence, hydrogen nucleation on the electrodes of electrolysis cells, flow around underwater propeller blades and jet break-up dynamics~\cite{Brennen2005}. In contrast to \textit{spinodal decomposition} where the initial state is thermodynamically unstable and phase separation occurs immediately in response to infinitesimal extensive perturbations~\cite{Debenedetti1996}, \textit{nucleation} is associated with the transition from a metastable state to a more stable one which can be far apart on the phase diagram. To phase-separate, a localized, finite-amplitude perturbation is required to overcome the energy barrier of forming a critical-size gas/liquid embryo. While in the more common heterogeneous nucleation, initial inhomogeneities are present due to natural impurities such as gas pockets and solid particles, in homogeneous nucleation~\cite{Lohse2016} the dynamics of nucleation is the result of the competition between thermal fluctuations and the energy barrier which is the characteristic of the metastable phase.

Several physical properties come into play influencing nucleation and cavitation, namely: (i) hydrodynamic thermal fluctuations, (ii) the surface tension between different phases and (iii) the cost of formation of a critical size gas/liquid embryo able to overcome the nucleation free-energy barrier. These properties are typically addressed in (i) the stochastic hydrodynamic approach~\cite{LANDAU1992}, (ii) in the thermodynamics of multi-phase interfaces~\cite{Rowlinson1983} and (iii) Classical Nucleation theory (CNT)~\cite{Debenedetti1996, Kalikmanov2013}. Computer simulations of nucleation are commonly conducted using the microscopic method molecular dynamics (MD) which can only treat a small domain with limited number of particles, where the role of hydrodynamics can be taken into account only at very high computational cost. Recently, homogeneous and heterogeneous nucleation in presence of fluctuating hydrodynamics has been considered in the context of finite-difference methods yielding interesting results~\cite{Chaudhri2014,Gallo2018} that showcased the technical advantages of mesoscopic models over MD approaches~\cite{Gallo2018}.  However, the characteristics of metastability, \textit{i.e.}, a local equilibrium that can withstand infinitesimal
perturbation, can only be addressed near the critical point.

The mesoscopic lattice Boltzmann method has achieved significant success in hydrodynamic simulations~\cite{kruger2017lattice, succi2018lattice}. In this paper we demonstrate the ability of Shan-Chen lattice Boltzmann model (LBM)~\cite{Shan1993,Shan1994a} to correctly capture the unstable and metastable dynamics through a coherent inclusion of stochastic hydrodynamics. In the unstable region of the phase diagram, immediate phase separation is observed for very small amplitude perturbations, and follows the same pattern observed in molecular dynamics simulations~\cite{Binder2012}, \textit{i.e.}, reaching in the final stage interfaces with different topology as a function of the initial density. In the metastable region phase-separation occurs on a time scale that diverges as the noise amplitude goes to zero. The system response to thermal fluctuations is studied by means of a stochastic hydrodynamics approach~\cite{LANDAU1992,Gross_2010,Gross_2011}. The structure factor (or form factor), $S(k)$, which converges to the isothermal compressibility in the long-wavelength limit, is obtained theoretically and verified numerically to a high precision. In particular, because of the mathematical structure of the model, an \emph{exact} expression for $S(k)$ is obtained. Two different scaling regimes of $S(k)$ are probed: (i) w.r.t.\ the temperature difference with critical point along the equilibrium binodal curve, and (ii) w.r.t.\ pressure difference with the spinodal values along an isotherm in the metastable region. In both cases the compressibility is found to diverge with exponents values in agreement with theoretical predictions~\cite{Debenedetti1996}.
\begin{figure}[!ht]
	\centering
	\includegraphics[scale=0.13]{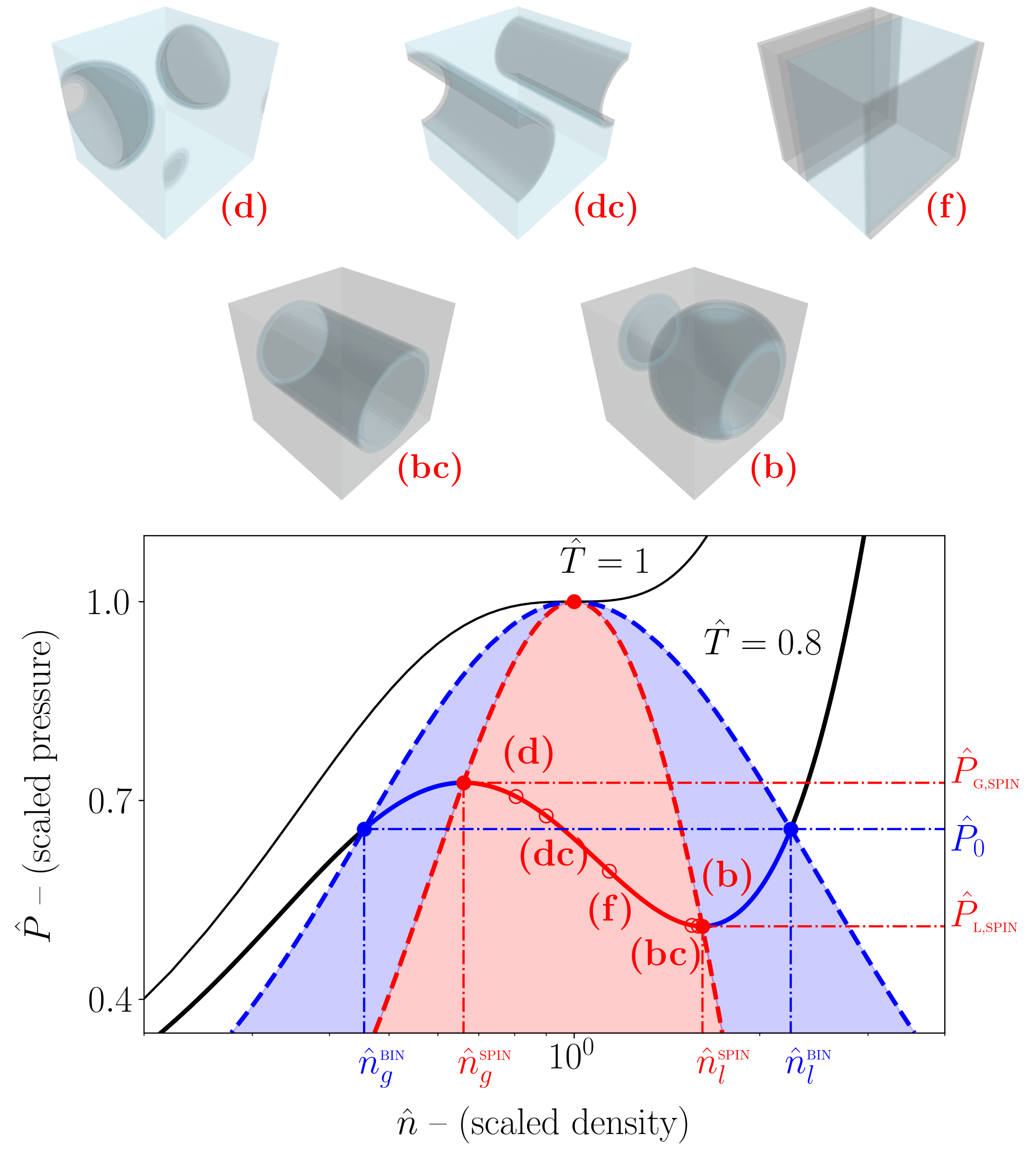}
	\caption{Lower panel: Isotherms at the critical temperature ($\hat{T} = 1$) and sub-critical temperature of $\hat{T} = 0.8$. Dashed blue and red lines are respectively the binodal and spinodal curves, and the blue- and red-shaded areas the metastable and unstable regions. The five points on the unstable part	of the isotherm (red solid line) correspond to the initial densities which phase-separate immediately with final interface topologies of droplet (d), droplet-cylinder (dc), flat (f), bubble-cylinder (bc) and bubble (b), all shown in upper panel.}
	\label{fig:sketch}
\end{figure}

LBM originated from a lattice-gas fluid model~\cite{kruger2017lattice,succi2018lattice} and later reformulated as a special velocity-space discretization of the Boltzmann equation.  The single-particle distribution function, $f(\bx, \bxi, t)$, is simplified to its values on a small set of discrete velocities while preserving the dynamics of its moments, and hence the hydrodynamics~\cite{Shan1998,Shan2006b}. The Shan-Chen non-ideal-gas model incorporates the inter-particle interaction through a mean-field Vlasov force
\begin{equation}
	\label{eq:SCForce}
	\bm{F}(\bx) = G\psi(\bx)\sum_{a=1}^{N_F} w(|\ea|^2)\psi(\bx+\ea)\ea,
\end{equation}
where $G>0$ is an interaction strength, $\{\ea: a = 1, \cdots, N_F\}$ the vectors pointing from $\bx$ to its interacting neighbors on the lattice, $w(|\ea|^2)$ a set of weights, and $\psi(\bx)$ the so-called \textit{pseudo-potential} encapsulating the details of interaction over fixed distances.  Together with the symmetry of the interacting set, the carefully chosen weights, $w(|\ea|^2)$, ensure macroscopic isotropy~\cite{Shan08, Sbragaglia2007,Lulli_2021}.

The interaction alters the equation of state (EoS) by adding a non-ideal-gas contribution to the pressure, yielding the following non-ideal-gas EoS:
\begin{equation}
	\label{eq:eos}
	p(n, T) = nT - \frac{G}2\psi^2(n),
\end{equation}
where $n$ and $T$ are respectively the number density and dimensionless temperature. Although the original model~\cite{Shan1993} was defined for an isothermal underlying LB model, the above EoS can be easily verified in thermal models.

We first show that the iso-thermal compressibility, $\kappa_T\equiv \hat{n}^{-1} \partial\hat{n}/\partial \hat{p}$, obtained from Eq.~(\ref{eq:eos}) diverges with the correct exponents~\cite{Debenedetti1996} as the system approaches from a stable configuration to the critical point ($\hat{T}\rightarrow 1$), or the spinodal curve ($\hat{n}\rightarrow\hat{n}_s$). In the present study, $\psi$ is set to $\exp(-1/n)$~\cite{Shan1994a}. Introducing two scaling constants, $a$ and $b$, such that $\psi = a\exp(-b/n)$, the EoS becomes 
\begin{equation}
  p = nT - \frac{Ga^2}2\exp\left(-\frac{2b}n\right).
\end{equation}
As $a^2$ also regulates the interaction strength, $G$ is omitted hereinafter to remove the redundancy.  Letting the critical density, temperature and pressure be denoted by $n_c$, $T_c$ and $p_c$ respectively and using the conditions $\partial p/\partial n = \partial^2 p/\partial n^2 = 0$ at the critical point, we can solve $a$ and $b$ in terms of $n_c$ and $T_c$ as
\begin{equation}
  a = e\sqrt{T_cn_c},\quad\mbox{and}\quad b = n_c.
\end{equation}
$\psi$ and the EoS are then expressed in critical quantities as:
\begin{align}
	\psi(n) & = \sqrt{T_cn_c}\exp\left(1-\frac{n_c}n\right),\\
	p & = nT - \frac{n_cT_c}2\exp\left(2-\frac{2n_c}n\right).\label{eq:pbulk}
\end{align}
The critical pressure is equal to $p_c \equiv p(n_c, T_c) = n_cT_c/2$. In terms of the \textit{reduced} quantities: $\hat{n} = n/n_c$, $\hat{p} = p/p_c$ and $\hat{T} = T/T_c$, the EoS becomes:
\begin{equation}
	\label{eq:PRescaled}
	\hat{p} = 2\hat{n}\hat{T} - \exp\left(2-2/\hat{n}\right).
\end{equation}
The rescaled EoS, known as the \textit{general EoS}, has identical shape as the unscaled one but a critical point at $\hat{n} = \hat{p} = \hat{T} = 1$. The phase diagram is shown in Fig.~\ref{fig:sketch}.

By taking the inverse of the derivative $\partial/\partial\hat{n}$ of Eq.~\eqref{eq:PRescaled}, we have
\begin{equation}
	\kappa_T = \frac{\hat{n}}2
	\left[\hat{n}^2\hat{T} - \exp\left(2 - \frac 2{\hat{n}}\right)\right]^{-1}.
\label{eq:k_crit_scaling}
\end{equation}
Since $\hat{n}\cong 1$ near critical point, $\kappa_T\sim \left(\hat{T} - 1\right)^{-1}$ as $\hat{T}\rightarrow 1$. Now consider a point, $\left(\hat{p}_s, \hat{n}_s\right)$, on the spinodal curve. Using the fact that $\partial\hat{p} / \partial{\hat{n}} = 0$, the leading-order expansion in the vicinity is
\begin{equation}
	\hat{p} - \hat{p}_s \cong\frac 12\left.\frac{\partial^2\hat{p}}{\partial \hat{n}^2}
	\right|_{\hat{n}=\hat{n}_s,\hat{T}}\left(\hat{n} - \hat{n}_s\right)^2.
\end{equation}
Denoting $A = \left.\partial^2\hat{p}/\partial \hat{n}^2 \right|_{\hat{n}=\hat{n}_s,\hat{T}}$ for brevity, we have
\begin{equation}
	\kappa_T \cong \frac 1{A\hat{n}_s}\left(\hat{n} - \hat{n}_s\right)^{-1}
	\cong\frac{1}{\sqrt{2A}\hat{n}_s}\left(\hat{p} - \hat{p}_s\right)^{-1/2}.
\label{eq:k_spin_scaling}
\end{equation}
Eqs~\eqref{eq:k_crit_scaling} and~\eqref{eq:k_spin_scaling} yield the critical and spinodal scaling of $\kappa_T$, respectively, which will be shown below to match the correlation function of the density fluctuations in the long-wavelength limit.

We now discuss the phase-separation as a response to thermal fluctuations. In the Landau hydrodynamic fluctuation theory~\cite{LANDAU1992}, thermal fluctuations are included into hydrodynamics by the `outside'~\cite{LANDAU1992} Langevin-type noise stress, $\bm{R}$
\begin{subequations}
	\label{eq:ns}
\begin{gather}
	\partial_t {n} + \nabla \cdot(n\bu) = 0,\\
	\label{eq:ns1}
	\partial_t (n\bu) + \nabla\cdot(n\bu\bu) = -\nabla\cdot\bm{P} + \nabla\cdot\bm{\sigma}
	+ \nabla\cdot\bm{R},\\
	\bm{\sigma} = \mu_s\left[\nabla\bu + \left(\nabla\bu\right)^T -\frac{2}{D}\bm{I}\nabla\cdot\bu\right] + \mu_b\bm{I}\nabla\cdot\bu.
\end{gather}
\end{subequations}
where $\bu$ is the fluid velocity, $\bm{\sigma}$ is the Newtonian deviatoric stress, $\bm{P}$ the pressure tensor containing both the hydrostatic term and the one due to inter-molecular interaction which vanishes in the ideal-gas limit, $\mu_s$, $\mu_b$ are the shear/bulk viscosity, respectively, and $D$ the space dimensionality. $\bm{R}$ is a spatially and temporally uncorrelated Gaussian random stress with variance proportional to the \emph{hydrodynamic fluctuations energy} $k_B\vartheta$ where $k_B\simeq1.380649\times 10^{-23}$ (J/K) (for more details see SM). We are working in the isothermal framework for which the background temperature is constant and fluctuations only enter through the stress tensor~\cite{Gross_2010, Gross_2011}. It would be very interesting to extend the present approach by including the heat current as already proposed in the general theory.
\begin{figure}[t]
	\centering
	\includegraphics[scale=0.3]{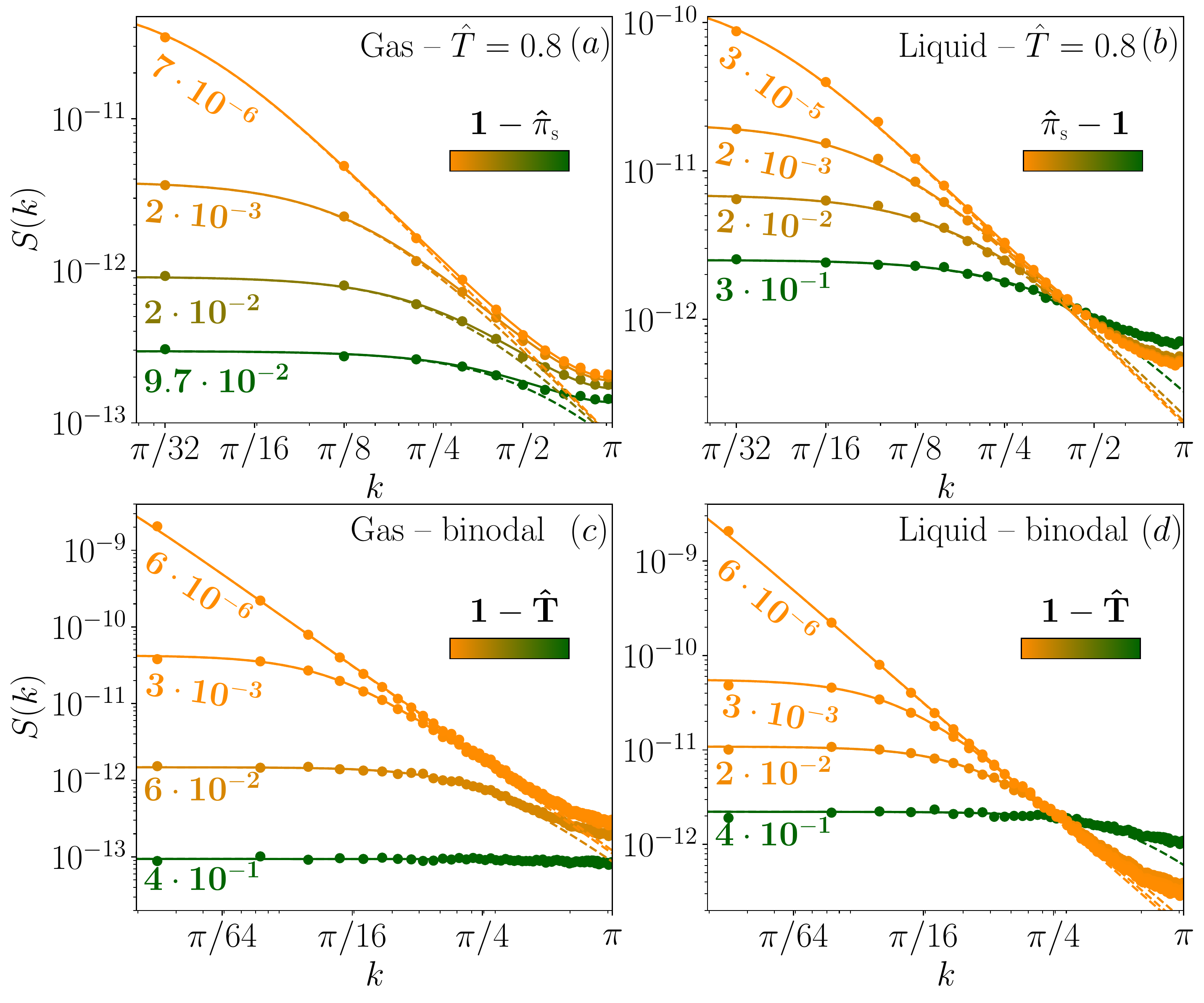}
	\caption{Density structure factor $S(k)$ as a function of $k$ at varying pressure along the $\hat{T} = 0.8$ isotherm (top row), and at varying temperature along the binodal (bottom row).  The left and right columns show the cases starting from the gas and liquid phasse respectively. The color scale, from green to orange, indicates the distance in pressure from the spinodal line or in temperature from the critical point. In dashed we report Eq.~\eqref{eq:sk_exact} truncated at $O(k^2)$ yielding a significant difference for $k\gtrsim\pi/2$.} \label{fig:sk_vs_k}
	%\label{fig:sk_spinodal_scaling}
\end{figure}
\begin{figure}
	\centering
	\includegraphics[scale=0.3]{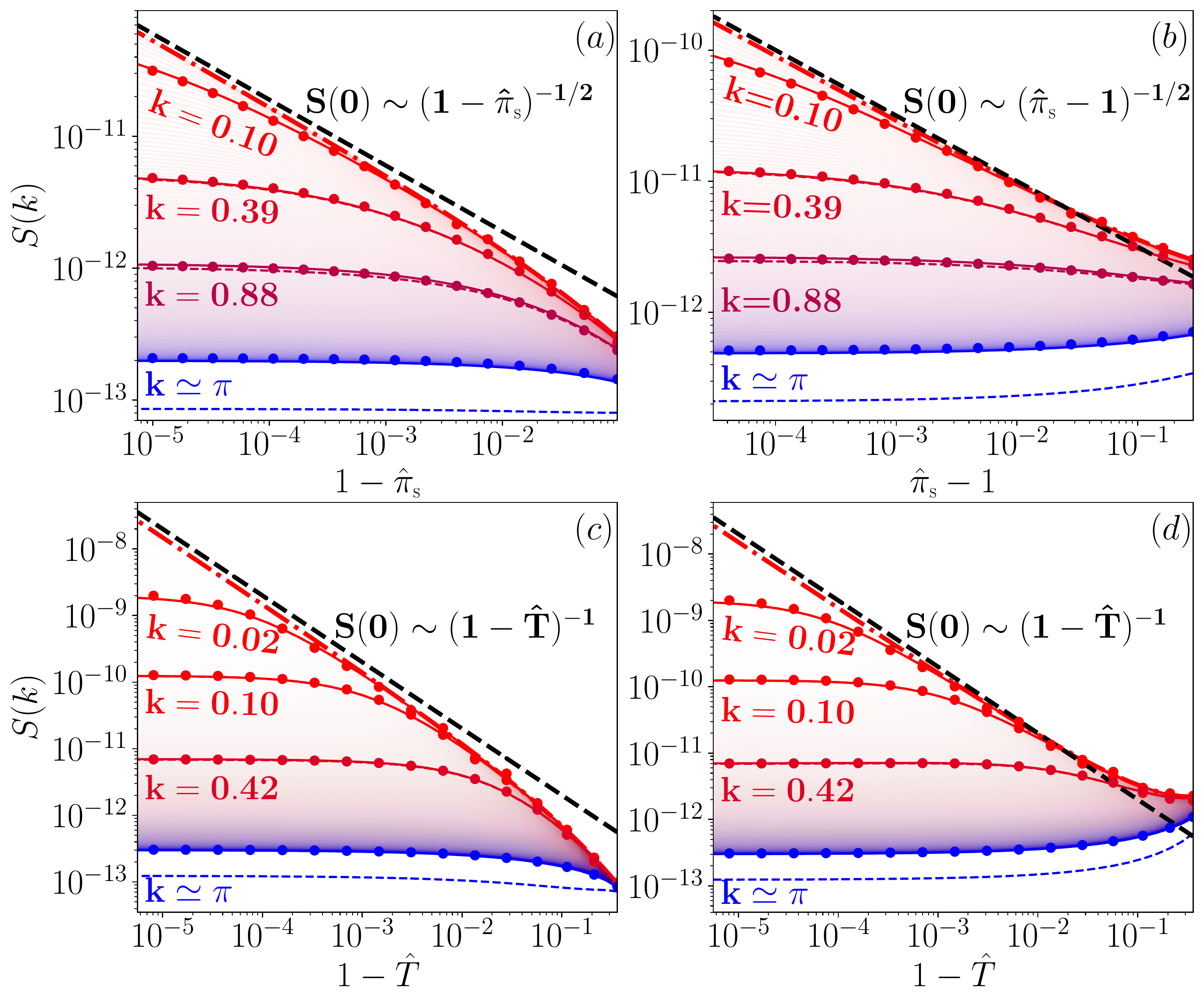}
	\caption{$S(k)$ as functions of the relative pressure from the spinodal line $|1-\hat{\pi}_{\text{s}}|$ (top), and relative distance from the critical point $1 -\hat{T}$ (bottom). Solid lines indicate the analytical prediction while the dashed lines the truncation to $O(k^2)$. Upon approaching the spinodal line or the critical point the large scale ($k\to0$ in dot-dashed red) converges to a power-law behavior (in dashed black).} \label{fig:sk_vs_dist}
 %\label{fig:sk_critical_scaling}
\end{figure}
A framework for including thermal fluctuation in LBM was developed by adding noise terms to the LBM equation~\cite{Gross_2010,Gross_2011}
\begin{equation}
	f_i(\bx+\bxi_i,t+1) - f_i(\bx, t) = \Omega_i(f_i) + F_i + \eta_i,
\end{equation}
where $\Omega$ is the usual Bhatnagar–Gross–Krook (BGK) collision term~\cite{kruger2017lattice, succi2018lattice} giving rise to the Navier-Stokes hydrodynamics. $\eta_i$ is the Langevin-type noise with a covariance matrix diagonal in the moment space assuring the local conservation
of the hydrodynamic moments~\cite{Gross_2010, Gross_2011}. $F_i$ is the forcing term~\cite{Guo2002} used to implement Eq.~\eqref{eq:SCForce}. We now turn our attention to the structure function $S\left(k\right)=\langle|\delta\hat{n}({k})|^{2}\rangle$, with the angle brackets $\langle \ldots \rangle$ indicating the steady state average. This is the Fourier transform of the density correlation function connected to the direct pair correlation function at the foundation of Ornstein-Zernike theory~\cite{Gross_2010, Gross_2011}. The structure factor can be directly measured in experiments and compared with simulations data~\cite{Thomas_2007}, hence a relevant quantity proportional to the isothermal compressibility $\kappa_T$ in the long-wavelength limit, i.e. $S(0)\sim \kappa_T$~\cite{Kalikmanov_2001_Book}. This allows to assess the robustness of the simulations by studying the two scaling limits discussed above, i.e. in temperature towards the critical and pressure towards the spinodal points. To calculate $S(k)$ we start from the Fourier transform of the Navier-Stokes equations~\cite{Gross_2010, Gross_2011} linearized around the quiescent homogeneous state $\langle n\rangle=n_0$ and $\bm{u} = \delta \bm{u}$
\begin{equation}
\begin{split}
\partial_{t}\delta\hat{n}&=\imath n_{0}k\delta\hat{u}_{||},\\
\partial_{t}\delta\hat{u}_{||}&=\frac{\imath k}{n_{0}}\left(\delta\hat{P}_{||}-\hat{R}_{||}\right)-\nu_{||}k^{2}\delta\hat{u}_{||},\\    
\end{split}
\label{eq:lin-nse}
\end{equation}
where $u_{||}=\bm{u}\cdot \hat{\mathbf{k}}$, $\nu_{||}=\nu_{\text{b}}+2\left(1-1/D\right)\nu_{\text{s}}$ is the longitudinal viscosity and $\hat{R}_{||}=\hat{\bm{R}} : \hat{\mathbf{k}}\hat{\mathbf{k}}$. The key quantity is $\delta\hat{P}_{||}=\delta\hat{\bm{P}}:\hat{\mathbf{k}}\hat{\mathbf{k}}=\bar{c}_{s}^{2}\left(k\right)\delta\hat{n}$, defining the $k$-dependent speed of sound~\cite{Gross_2010, Gross_2011} $\bar{c}_{s}^{2}(k)$. For the SC model there exist a lattice expression for the pressure tensor which enjoys an important property for flat interfaces: the normal component is constant to machine precision throughout the interface thus realizing the mechanical equilibrium condition $P_{\text{N}}=P_0$ \emph{on the lattice}~\cite{Shan08,Lulli_2021}. One can notice that the variation of the longitudinal component of the pressure tensor is actually proportional to $P_{\text{N}}$ given that $\hat{\mathbf{k}}$ is in the direction of the density gradient, hence one can leverage the one-dimensional expression
\begin{equation}
P_{\text{N}}\left(x\right)=n\left(x\right)T-\frac{1}{4}\psi\left(x\right)\left[\psi\left(x+1\right)+\psi\left(x-1\right)\right].
\label{eq:PN_lattice}
\end{equation}
As detailed in the SM one can take the variation of~\eqref{eq:PN_lattice} as $\delta P_{\text{N}}=\mbox{d} P_{\text{N}}/\mbox{d} n|_{n_0} \delta n$, take its Taylor expansion in real space and finally its Fourier transform. By virtue of the definition of $P_{\text{N}}$ it is possible to sum the series in $k$ and obtain for the scale-dependent speed of sound
\begin{equation}
\bar{c}_{s}^{2}\left(k\right)=\bar{c}_{s,0}^{2}-\frac{1}{2}\psi_{0}\psi_{0}'\left[\cos\left(k\right)-1\right],
\end{equation}
where $\bar{c}_{s,0}^{2} = \mbox{d}p/\mbox{d}n$, where $p$ is defined in Eq.~\eqref{eq:pbulk}. This result simply follows from the fact that Eq.~\eqref{eq:PN_lattice} is closely related to the lattice Laplacian~\cite{GrossThesis, parisi1998statistical}. It is possible to combine the two Eqs~\eqref{eq:lin-nse} and obtain a second order equation in time for $\delta \hat{n}$, which can be again Fourier transformed in the frequency domain and yield an algebraic expression for $\delta n(\bm{k},\omega)$. Hence, one defines the dynamic structure factor $S\left(\mathbf{k},\omega\right)=\langle|\delta\hat{n}\left(\mathbf{k},\omega\right)|^{2}\rangle$. The frequency dependence can be integrated out by considering a complex contour integral around the poles yielding $S\left(k\right)=n_{0}k_B\vartheta/{\bar{c}_{s}^{2}\left(k\right)}$, i.e.
\begin{equation}
S\left(k\right)=n_{0}k_B\vartheta\Big/\left\{ \bar{c}_{s,0}^{2}-\frac{1}{2}\psi_{0}\psi_{0}'\left[\cos\left(k\right)-1\right]\right\},
\label{eq:sk_exact}
\end{equation}
where $\psi_0 = \psi(n_0)$, $\psi' = \mbox{d}\psi/\mbox{d}n$. We remark that this expression is an exact identity valid for all values of $k$ and $\hat{T}$ in contrast to the previous approaches~\cite{Gross_2010, Gross_2011, Chaudhri2014, Gallo2018} which are limited to $O(k^2)$ and only reliable near the critical point. Finally we remark the similarity of Eq.~\eqref{eq:sk_exact} to the momentum space propagator of the lattice Gaussian model~\cite{parisi1998statistical} (see SM for details).

To include hydrodynamic fluctuations in simulation, we extended~\cite{Gross_2010, Gross_2011} to three dimensions (3D). Key to the implementation is a proper choice of the covariance matrix for the noise populations $\eta_i$. As observed in~\cite{Gross_2011}, one only needs to consider a diagonal correlation matrix in the space of the noise hydrodynamic moments $N_a$ which are obtained by some specific linear combinations of the noise populations, i.e. $N_a = \sum_i m_{ai} \eta_i$, hence $\Xi_{ab}=\langle N_a N_b\rangle \propto k_B\vartheta \delta_{ab}$ with $a>3$. The covariance of the first four moments, $a=0, 1, 2, 3$, are set to zero to ensure mass and momentum conservation.

We performed two kinds of simulations in periodic 3D domains of linear size $L$ with homogeneous density $n_0$ as initial condition: (i) $n_0$ is chosen along the isotherm $\hat{T}=0.8$ in the unstable region with the fluctuations energy set to $k_B\vartheta=10^{-10}$ (lbu)~\footnote{lbu stands for Lattice Boltzmann units, given that we are not matching a specific physical system. The order of magnitude of $k_B \vartheta$ is compatible with the previous LBM literature~\cite{Gross_2010, Gross_2011}}, and (ii) $n_0$ belongs to either the liquid or the gas phases along (a) the isotherm $\hat{T}=0.8$ in the metastable region approaching the spinodal points or (b) along the binodal approaching the critical point while keeping $k_B\vartheta=10^{-13}$ (lbu) in order to avoid nucleation. In the upper panels of Fig.~\ref{fig:sketch} we report the final density configurations in the unstable region: Our simulations display qualitative the same sequence of interface shapes as in~\cite{Troster2011}. Indeed, the MD simulations and the present LBM approach share the main symmetry feature of exact mass conservation. This results can be used to estimate the curvature dependence of the surface tension~\cite{Troster2011}: The connection with the recently proposed SC-LBM approach for the estimation of the Tolman length~\cite{Tolman1949, Lulli_2022} will be the subject of a future work.

Shown in Figs.~\ref{fig:sk_vs_k} and~\ref{fig:sk_vs_dist} are the numerically measured structure factor together with the values of Eq.~\eqref{eq:sk_exact}. We note that the finite simulation domains of $L=128$ and $L=512$ impose an artificial minimum wave number of approximately $10^{-3}$. A number of interesting aspects are to be seen. First, in contrast to previous results~\cite{Gross_2011}, numerical measurements are in excellent agreement with the  exact expression across several decades of wave length and distance from the critical and spinodal line. For comparison, Eq.~\eqref{eq:sk_exact} truncated to $O(k^2)$ is also shown as the thin dashed lines: the agreement worsens for $k\gtrsim \pi/2$. Secondly, the scaling at both $\hat{\pi}_s=\hat{P}/\hat{P}_s\rightarrow 1$ and $\hat{T}\rightarrow 1$ in Fig.~\ref{fig:sk_vs_k} and~\ref{fig:sk_vs_dist} indicate the diverging trends of $\kappa_T$ as $k\rightarrow 0$, since $S(0)\sim \kappa_T$~\cite{Kalikmanov_2001_Book}. Thirdly, as can be read off the graphs in Fig.~\ref{fig:sk_vs_dist}, the structure factor exhibits the correct critical exponents as the system approaches either the spinodal curve or the critical point.

In summary, we studied in this Letter the metastable and unstable characteristics of the stochastic pseudo-potential lattice Boltzmann model in response to thermal fluctuation.  Using previously obtained pressure tensor, the structure factor is theoretically obtained and numerically verified to high precision. The metastable equilibriums are numerically confirmed as stable to small perturbations. Theoretically and numerically, the long-wave-length limit of the structure factor corresponding to the isothermal compressibility is found to diverge with the correct exponents as the system approaches to the spinodal curve or critical point.  With the metastable and unstable  characteristics quantitatively obtained, the present study paved the way for the pseudo-potential LB model to be used in modeling nucleation with hydrodynamic influences. All simulations have been run on an architecture-independent GPU/CPU implementation which can be found on the GitHub repository \href{https://github.com/lullimat/idea.deploy}{https://github.com/lullimat/idea.deploy}~\cite{pycuda_opencl}.

This work was supported by National Science Foundation of China Grants~12050410244 and 92152107, by Department of Science and Technology of Guangdong Province Grant 2020B1212030001, Shenzhen Science and Technology Program Grant KQTD20180411143441009, and from the European Research Council (ERC) under the European Union’s Horizon 2020 research and innovation programme (grant agreement No 882340).

\end{document}